\documentclass[twoside]{romjist}
\usepackage[top=4.3cm, bottom=5.7cm, left = 3.74cm, right = 3.75cm]{geometry}
\usepackage{times}
\usepackage{mathptmx}
\usepackage{authblk}
\usepackage{graphicx} 
\usepackage{amsmath}
\usepackage{amssymb}
\usepackage{amsthm}
\usepackage{cite}
\usepackage{graphicx}
\usepackage{epstopdf}
\usepackage{amssymb}
\usepackage{centernot} 
\usepackage{float}
\usepackage{multirow} 
\usepackage{footnote}
\usepackage{mathtools}

\usepackage{enumitem}
\usepackage{slashbox}
\makesavenoteenv{tabular}
\makesavenoteenv{table}

\begin{document}
    
\issuevolume{\textbf{22}}
\issuenumber{1}
\issueyear{2019}
\issuepages{3--13}

\headauthors{T. Sandu \textit{et al.} }
\headtitle{\textit{Modal Approach to the Theory of Energy Transfer Mediated by a Metallic Nanosphere}}

\setcounter{page}{1}

\title{Modal Approach to the Theory of Energy Transfer Mediated by a Metallic Nanosphere} 

\author[ ]{Titus SANDU}
\author[ ]{Catalin TIBEICA}
\author[ ]{Oana T. NEDELCU}
\author[ ]{Mihai GOLOGANU}

\affil[ ]{\small{National Institute for Research and Development in Microtechnologies-IMT, 126A, Erou Iancu Nicolae Street, Bucharest, Romania}}

\affil[ ]{\texttt{E-mail: titus.sandu@imt.ro}}

\maketitle
    
\begin{abstract}
 Theoretically, the presence of a metallic nanoparticle enhances the
 intermolecular energy transfer. We calculate this enhancement factor
 with a modal approach pertaining analytical results in the case of a
 nanosphere. We calculate the Green's function of the system relaying
 on the spectral properties of the electrostatic operator, fully known
 for spherical geometry. In contrast to other treatments, the present
 calculations are straightforward for any molecular orientation giving
 modal information about the response of the system. Numerical
 calculations and further discussions are also provided.
 
\end{abstract}


\section{Introduction}

\hspace{0.5cm} Energy transfer, especially the resonance energy
transfer (RET) between a donor molecule and a near by acceptor
molecule, plays an important role in many aspects of photophysical and
photochemical processes like photosynthesis and light harvesting
[1,2], photovoltaics [3], biomolecular structure and fluorescence
probing [1,4-6], biosensing [7]. The main transfer mechanism of these
processes obviously depends on the distance between the donors and the
acceptors. Thus, when the distance is less than 2 nm the mechanism
depends on the molecular orbital overlap hence, the treatment is quantum
mechanical [2, 8]. Moreover, within this range of distances the energy
transfer enters the competition with the electron transfer process.
On the other hand, when distances are between 2 and 10 nm the
electrostatic dipole-dipole interaction enables a nonradiative energy
transfer by the Förster resonance energy transfer (FRET) [2,
  9]. Alternatively, the radiative mechanism plays a role when the
distances between donors and acceptors are comparable or greater than
the wavelength of the incoming radiation [10].
      
FRET is a weak process because it is inverse proportional to the sixth
power of the separation distance. Hence, an enhancement is often
desired in various applications like in the long-range energy transfer
between fluorescent centers [11].  The enhancement is obtained by
plasmonic effect through the coupling between light and the collective
excitations of the free electrons in metals. Plasmonic effects of
surface plasmon polaritons (occurring on the interfaces between metals
and dielectrics) and localized surface plasmons (around metallic
nanoparticles) are shown by strong fields and field confinements below
diffraction limit which can be utilized in sensing and biosensing
applications [12]. Strong fields lead to strong coupling of light with
atomic and atom-like systems [13-15].
       
FRET enhancement has been studied in planar metallic structures [16]
and in variously shaped nanoparticles: spheres and spheroids [17],
nanodiscs [18] and nanorods [19]. When it is possible analytical
treatments of a physical process can provide valuable insights of the
process. Analytical approaches of plasmonic enhancement of FRET were
used in [17] for spheres and spheroids, for shelled spheres in [20]
and for spheroids in [21]. For example Shishodia et al. used the
Bergman's approach [22], which has the disadvantage of operators
defined in the whole space [20].

In our work we use a method closely related [23-26]. It is also an
operator method, but the operators are defined on surfaces rather than
the whole space. Our method allows the calculation of a Green's
function that permits easy calculations the system response to any
kind of stimulus [27]. The Green's function exhibits a modal
decomposition of the response and it has been used in spectroscopies
like electron energy loss spectroscopy (EELS) or scanning near-field
optical microscopy (SNOM), but it hasn't been used in FRET problems
[27].  In the present work we apply the Green's function method to
estimate the FRET enhancement. The approach is based on the fact that
for spherical nanoparticles all spectral properties of the surface
operators are known [28], hence an analytical expression of the
boundary Green's function is provided.  It will be further shown that
the boundary Green's function allows easy calculations for any
orientation of donors and acceptors. In addition to that, the terms
associated with plasmonic enhancement can be easily identified in the
expressions of the FRET enhancement factor. In a recent conference
paper [29] we tackled the same problem. However, the approach is a bit
different here, mostly by setting the problem in a broader context and
explicitly calculating the boundary Green's function. The paper has
the following structure: in section 2 we present the calculation model
for plasmonic enhancement of FRET; in section 3 we present the
description of plasmonic enhancement of FRET by the boundary Green's
function, which allows a modal decomposition; in section 4, for a
nanosphere we calculate its boundary Green's function and its
plasmonic enhancement of FRET; in the last section, section 5, we
present some numerical results, the discussions, and the concluding
remaks.

\section{The model for plasmonic enhancement of FRET}
\hspace{0.5cm} Energy transfer by FRET is governed by the
electrostatic dipole-dipole interaction, which provides an expression
for rate transfer of the following form [2, 9, 10, 30]:

\begin{equation} \label{eq1} 
  k_{ET}=\frac{\eta_D}{\tau_D}
  \frac{9000 k^2 \ln 10}{128\pi^5 n^4 N_A }\frac{1}{R^6}\int
  \frac{\varepsilon(v)I(v)}{v^4} d v .
\end{equation}

\noindent In Eq.~(\ref{eq1}) $\eta_D$ and $\tau_D$ are respectively
fluorescence quantum yield of the donor and donor emission lifetime,
$k$ is an orientation factor of the dipoles associated with the donor
and the acceptor, $N_A$ is the Avogadro number, $n$ is the refraction
index of the host medium, $R$ is separation distance between donor and
acceptor, $v$ is wavenumber, while the integral is the spectral
overlap between absorption coefficient of the acceptor $\varepsilon
(v)$ and the normalized emission spectrum of the donor
$I(v)$. Nonetheless, the F\"oster theory based on Eq. (1) has several
shortcomings, a few of which were mentioned above, like the range of
validity (between 2 and 10 nm). Below 2 nm the point-dipole
approximation is not valid anymore; also other quantum mechanical
processes may become relevant like the electron transfer between donor
and acceptor [2,8,10]. For distances comparable with the light
wavelength radiative processes are needed to be considered [10,30]. In
addition, the quantum nature of light may be also considered. Moreover,
FRET is based on weak coupling and index of refraction may be
inhomogeneous. Many of these limitations as well as the plasmon
enhanced FRET have been considered in recent works where the quantum
nature of the donors, acceptors, and radiation has been considered
[30]. The expression reads:
                      
\begin{equation} \label{eq2} 
  k_{QET}=\frac{\eta_D}{\tau_D} \frac{9000 \ln10}{128\pi^5 N_A}\int
  \frac{\varepsilon(v)I(v)}{v^4} \Bigg| \frac{{\mathbf e}_A \cdot
    \mathbf{E}^D({\mathbf r}_A,v)}{d_D} \Bigg|^2 dv.
\end{equation}

\noindent In the last term of right hand side, $\mathbf{e}_A$ is the
direction of the acceptor dipole, located at $\mathbf{r}_A$,
$\mathbf{E}^D$ is the field created by the donor with a classical
dipole strength $d_D$ at acceptor site. The information about dipole
orientation factor, distance dependence as well as about the
refraction index of the medium is embedded in this term. However, the
plasmon enhanced factor is still calculated by classical
electrodynamics means.

Below it is provided the method used in the electrostatic
approximation by which we can calculate the plasmonic enhancement
factor of FRET. Thus both the donor (D) and the acceptor (A) are
associated with point-like dipoles, $\mathbf{d}_D$ and $\mathbf{d}_A$,
respectively, whose the energy transfer is governed by the
dipole-dipole interaction [2,9 ,10]. The energy transfer and
eventually its enhancement regard D and A as time-harmonic point-like
dipoles interacting with each other directly or via the metallic
nanoparticle (Fig. 1) [17,21]. The electrostatic potential contains
four terms:
       
\begin{equation} \label{eq3} 
  \Phi(\mathbf{r})=\Phi_A(\mathbf{r})+\Phi_D(\mathbf{r})+\Phi_{A_{ind}}(\mathbf{r})+\Phi_{D_{ind}}(\mathbf{r}).
\end{equation}

The first and the second term are the electric potentials generated by
A and D, while the other two terms are the electric potentials
generated by the charge induced on the nanoparticle by the two dipoles
A and D. The electric field at the acceptor site is:
       
\begin{equation} \label{eq4} 
  \mathbf{E}_A=-\nabla \big[\Phi_D(\mathbf{r}_A)+\Phi_{D_{ind}}(\mathbf{r}_A)+\Phi_{A_{ind}}(\mathbf{r}_A)\big],
\end{equation}    
       
\noindent which gives an interaction energy for the acceptor in the
presence of both the donor and the nanoparticle of the following form:
       
\begin{equation} \label{eq5} 
  U_A=-\mathbf{d}_A \cdot \mathbf{E}_A=U_{AD}+U_{AD_{ind}}+U_{AA_{ind}}.
\end{equation}     
       
The plasmonic enhancement factor of the FRET process is thus defined
simply as [17,21]:
       
\begin{equation} \label{eq6} 
  \big|A\big|^2=\Bigg|1+\frac{U_{AD_{ind}}}{U_{AD}}\Bigg|^2.
\end{equation}

\section{Description of the plasmonic enhancement of FRET by the boundary Green's function}

\hspace{0.5cm} In the electrostatic (nonretarded) approximation the
total electric field $\mathbf{E}$ is the sum of the electric field $\mathbf{E}_{\textit free}$
of the imposed free charge $\rho_{\textit{free}}(\mathbf{r}, \omega)$ and the field of
the induced charge due to the presence of the nanoparticle
$\mathbf{E}_{\textit bound}$. It is assumed that the free charge distribution is placed
in a medium of dielectric permittivity $\epsilon_0$, outside the
nanoparticle of dielectric permittivity $\epsilon_1$, and has temporal
evolution of the form $exp(j\omega t)$. Accordingly all the electric
fields $\mathbf{E}_{\textit free}$, $\mathbf{E}_{\textit bound}$, and $\mathbf{E}$ have an associated
electrostatic potential, \textit{i. e.},
$\mathbf{E}_{\textit free}=-\nabla\Phi_{\textit free}$, $\mathbf{E}_{\textit bound}=-\nabla\Phi_{\textit bound}$, and the total electric field
$\mathbf{E}=-\nabla\Phi$, such that the electrostatic potential in the presence
of the nanoparticle has the form [23-27]:
                                
\begin{equation} \label{eq7} 
  \Phi(\mathbf{r},\omega)=\Phi_{\textit free}(\mathbf{r},\omega)+\Phi_{\textit bound}(\mathbf{r},\omega)
\end{equation} 
                
\noindent  with             

\begin{equation} \label{eq8} 
  \Phi_{\textit free}(\mathbf{r},\omega)= \int\frac{1}{\epsilon_0}G_{\textit free}(\mathbf{r},\mathbf{r}')\rho_{\textit free}(\mathbf{r}',\omega)d\mathbf{r}'
\end{equation}    

\noindent  and           

\begin{equation} \label{eq9} 
  \Phi_{\textit free}(\mathbf{r},\omega)= \int\limits_\Sigma G_{\textit free}(\mathbf{r},\mathbf{r}')\sigma(\mathbf{r}',\omega)d\Sigma_{r'}.
\end{equation}

The first integral is a volume integral in whole space, while the
second integral is on the surface $\Sigma$ that bounds the
nanoparticle. The charge density $\sigma(\mathbf{r},\omega)$ induced
on the surface of the nanoparticle is [23-26]:
  
\begin{equation} \label{eq10} 
  \sigma(\mathbf{r},\omega)=
  \sum\limits_k\frac{1}{\frac{1}{2\lambda}-\chi_k}u_k(\mathbf{r})\langle
  v_k|\mathbf{n \cdot E}_{\textit free} \rangle,
\end{equation}

\noindent where $\lambda=(\epsilon_1-\epsilon_0) /
(\epsilon_1+\epsilon_0)$ with $\epsilon_1$ and $\epsilon_0$ being
eventually $\omega$-dependent, $u_k$ and $\chi_k$ are the
eigenfunctions and the eigenvalues of the electrostatic operator
defined on $\Sigma$ of the nanoparticle as [23-27]

\begin{equation} \label{eq11} 
  \hat{M}[\sigma]=-\int\limits_{\mathbf{r}' \in \Sigma}
  \sigma(\mathbf{r}')\frac{\partial}{\partial n_r}G_{\textit free}(\mathbf{r},
  \mathbf{r}')d\Sigma_{r'}
\end{equation}

\noindent and $ \langle v_k|{\mathbf n \cdot E}_{\textit free} \rangle $ are
the scalar products between the corresponding eigenfunctions $v_k$
of the adjoint operator $M^\dagger$ and the dot product $\mathbf {n \cdot   E}_{\textit free}$, 
with $\mathbf n$ the normal to the surface $\Sigma$ of the nanoparticle. In
Eqs. (8), (9), and (11)

\begin{equation} \label{eq12} 
  G_{\textit free}(\mathbf{r},\mathbf{r}')=\frac{1}{4\pi}\frac{1}{|\mathbf{r}-\mathbf{r}'|}
\end{equation}    

\noindent  is the free-space Green's function. It is well known [23-26] that  $u_k$ and $v_k$  are bi-orthonormal, \textit{i.e.}, $ \langle v_k|u_j \rangle = \delta _{ij} $ and the $v_k$  are obtained from  $u_k$ by       
          
\begin{equation} \label{eq13} 
  v_k(\mathbf{r})=\int\limits_{\mathbf{r}' \in \Sigma}u_k(\mathbf{r}')G_{\textit free}(\mathbf{r},\mathbf{r}')d\Sigma_{r'}.
\end{equation}

\noindent We can define a new Green's function $G_{\textit
  bound}(\mathbf{r},\mathbf{r}',\omega)$ , which is a boundary Green's
function that describes the response of the system to imposed free
charge $\rho_{\textit free}(\mathbf{r},\omega)$, such that the total
electric potential is then given by [23-27]

\begin{equation} \label{eq14} 
  \Phi(\mathbf{r}, \omega)=\int \frac{1}{\epsilon_0}G_{\textit
    free}(\mathbf{r},\mathbf{r}')\rho_{\textit
    free}(\mathbf{r}',\omega)d\mathbf{r}'+\int G_{\textit
    bound}(\mathbf{r},\mathbf{r}',\omega)\rho_{\textit
    free}(\mathbf{r}',\omega)d\mathbf{r}'.
\end{equation}     
          
\noindent Combining Eqs. (7), (9), (10), (12), (13), and (14) one can
obtain the expression of boundary Green's function as [27]

\begin{equation} \label{eq15} 
  G_{\textit
    bound}(\mathbf{r},\mathbf{r}',\omega)=\frac{1}{\epsilon_0}\sum\limits_k
  \frac{1}{\frac{1}{2 \lambda} - \chi_k}\phi_k(\mathbf{r})
  \tilde{\phi}_k(\mathbf{r}'),
\end{equation}        
          
\noindent   where      

\begin{equation} \label{eq16} 
  \phi_k(\mathbf{r}')= \int\limits_{\mathbf{r}'\in \Sigma}
  u_k(\mathbf{r}')G_{\textit free}(\mathbf{r},\mathbf{r}')d\Sigma_{r'}
\end{equation}

\noindent is the single-layer potential generated by the charge
density $u_k$ and
    
\begin{equation} \label{eq17} 
    \tilde{\phi}_k(\mathbf{r})= \int\limits_{\mathbf{r}'\in \Sigma}
    v_k(\mathbf{r}')\frac{\partial}{\partial
      \mathbf{n}_{r'}}G_{\textit
      free}(\mathbf{r},\mathbf{r}')d\Sigma_{r'}
    \end{equation}
    
\noindent is the double-layer potential of the dipole density
$v_k$. We note here that $G_{\textit
  bound}(\mathbf{r},\mathbf{r}',\omega)$ provides a modal
decomposition of the response. Moreover, it can take a manifestly
symmetric form in $\mathbf{r}$ and $\mathbf{r}'$, in which $\tilde{\phi}_k$ is replaced
by $\phi_k$ [27], however the calculation of the boundary Green's
function in its manifestly symmetric form requires the proper
normalization of $u_k$ by their bi-orthogonality with $v_k$.
    
\section{The boundary Green's function of a nanosphere and the calculation of plasmonic enhancement of FRET }

\hspace{0.5cm} In order to calculate the boundary Green's function one
needs the knowledge of spectral properties of the electrostatic
operator (11) and of its adjoint, namely $\chi_k$, $u_k$, and
$v_k$. There are several shapes having the associated electrostatic
operators with well established spectral properties [28]. One of these
shapes is sphere, whose free-space Green's function (12) has a
separated form in spherical coordinates given by

\begin{equation} \label{eq18} 
  G_{\textit free}(\mathbf{r},\mathbf{r}') =
  \sum_{l,m}\frac{1}{2l+1}\frac{r_{<}^l}{r_{>}^{l+1}}Y_{lm}(\theta,
  \varphi),
\end{equation}

\noindent where $Y_{lm}(\theta, \varphi)$ are the spherical harmonics
and $r_{<}$ and $r_{>}$ represent the minimum and respectively the
maximum of the pair $(r, r')$. The form given by Eq. (18) is a
well-known expression found in any textbook treating classical
electrodynamics [31]. What has been less known is the fact that this
separated form allows the calculations of both the eigenfunctions and
the eigenvalues of (11) for spherical shape [28]. Although for sphere
the electrostatic operator is symmetric, the eigenfunctions $u_k$, and
$v_k$ are not identical (they differ by a constant) since they are
related by Eq. (13). Thus, for a sphere of radius $a$ one has
      
\begin{equation} \label{eq19} 
  u_{lm}(\theta, \varphi)=\sqrt{\frac{2l+1}{a^3}}Y_{lm}(\theta,
  \varphi),
\end{equation}
      
\begin{equation} \label{eq20} 
  v_{lm}(\theta, \varphi)=\sqrt{\frac{1}{(2l+1)a}}Y_{lm}(\theta, \varphi),
\end{equation}
      
\noindent   and  
      
\begin{equation} \label{eq21} 
  \chi_l=\frac{1}{2(2l+1)}.
\end{equation}
Now it is straightforward to calculate the boundary Green's function
for a spherical particle by combining Eqs. (15)-(21). Its expression
is given by

\begin{equation} \label{eq22} 
  G_{\textit bound}(\mathbf{r},\mathbf{r}',\omega) =
  \frac{1}{\epsilon_0}\sum_{l,m}\frac{\epsilon_1-\epsilon_0}{l \epsilon_1+(l+1)\epsilon_0}\,\frac{l}{2l+1}\,\frac{a^{2l+1}}{r^{l+1}r'^{l+1}}Y_{lm}(\theta,
  \varphi)Y_{lm}^*(\theta', \varphi').
\end{equation} 
We easily see that the boundary Green's function is symmetric in $r$
and $r'$, even though Eq. (15) does not show it manifestly.
     
Formally, the charge density of a dipole of strength $\mathbf d$ located at
$\mathbf r'$ is $\rho_d(\mathbf {r})=\mathbf {d} \cdot \nabla_{\mathbf{r}'}\delta(r-r')$. Now we are able
to find the electric potential generated by a dipole $\mathbf {d}_D$ located at
$\mathbf {r}_D$ due to the presence of a sphere of radius a centered at origin
as

\begin{equation} \label{eq23} 
 \Phi_{D_{ind}}(\mathbf{r},\omega)=\frac{-1}{\epsilon_0}\sum_{l,m}\frac{\epsilon_1-\epsilon_0}{l \epsilon_1 + (l+1)\epsilon_0}\frac{l}{2l+1}\frac{a^{2l+1}}{r^{l+1}}Y_{lm}(\theta, \varphi)\mathbf{d}_D \cdot \nabla_{\mathbf{r}'}\Big(\frac{Y_{lm}^*(\theta', \varphi')}{r'^{l+1}}\Big) \Bigg |_{r'=r_d} .
\end{equation} 
Finally the interaction energy between a donor dipole $\mathbf {d}_D$ located at
$\mathbf {r}_D$ and an acceptor dipole $\mathbf {d}_A$ located at $\mathbf {r}_A$ due to the
presence of a sphere is given by the following

%

\begin{equation} \label{eq24} 
    U_{ADind}=\frac{1}{\epsilon_0}\sum_{l,m}\frac{\epsilon_0-\epsilon_1}{l \epsilon_1 + (l+1)\epsilon_0}\frac{l}{2l+1}\frac{a^{2l+1}}{2l+1}\mathbf{d}_A \cdot \nabla_{\mathbf{r}}\Big(\frac{Y_{lm}(\theta', \varphi')}{r^{l+1}}\Big) \Bigg |_{r=r_a} \hspace{-1pt} \mathbf{d}_D \cdot \nabla_{\mathbf{r}'}\Big(\frac{Y_{lm}^*(\theta', \varphi')}{r'^{l+1}}\Big) \Bigg |_{r'=r_d}  \hspace{-1pt} .
\end{equation} 

Equation (24) applies for any locations and orientations of donor and
acceptor molecules and as far as we know it has not been encountered
in the literature.  In the following we apply expression (24) to the
dipole settings shown in Fig. 1.

\begin{figure}[H]
  \centering
  \includegraphics[width=0.5\linewidth]{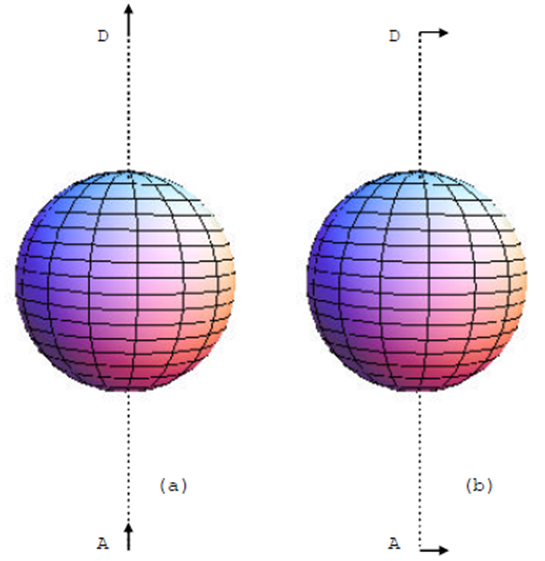}
  \caption{Arrangement geometries of the donor (D) and the acceptor
    (A) in the close proximity of a spherical metallic nanoparticle:
    (a) aligned and normal to surface dipoles, and (b) parallel and
    tangent to surface dipoles.}
  \label{fig:figure1}
\end{figure}

\section{Numerical results, discussions, and concluding remaks}

\hspace{0.5cm} The donors (D) in Fig. 1 have spherical coordinates
$r=r_D,\ \theta=0,\ \varphi=0$, while the acceptors (A) have the
coordinates $r=r_A,\ \theta=\pi$, and $\varphi=0$. We recall that the
gradient in spherical coordinates has the following expression: $
\nabla=\mathbf{e}_r\frac{\partial}{\partial_r}+ \mathbf{e}_{\theta}\frac{1}{r}\frac{\partial}{\partial_{\theta}} + 
\mathbf{e}_{\varphi}\frac{1}{r\,sin(\theta)}\frac{\partial}{\partial_{\varphi}}
$, where $\mathbf{e}_r$, $\mathbf{e}_\theta$ and $\mathbf{e}_\varphi$ are the local and
orthogonal unit vectors [31]. Moreover, the donor dipole is oriented
along $\mathbf{e}_r$ in Fig. 1(a) and along $\mathbf{e}_\theta$ in Fig. 1(b). Similarly
the acceptor is aligned with $-\mathbf{e}_r$ in Fig.1(a) and aligned with
$-\mathbf{e}_\theta$ in Fig. 1(b).  Now direct calculations lead us to the
plasmonic enhancement factor of the FRET process for dipole
arrangements shown in Fig, 1(a) of the form [20, 29]

\begin{equation} \label{eq25} 
  |A|^2=\Bigg|1+\frac{(r_A+r_D)^2}{2a^3}\sum_l (-1)^{l+1}\frac{l (l+1)^2(\epsilon_1-\epsilon_0)}{l \epsilon_1+ (l+1)\epsilon_0}\Bigg( \frac{a^2}{r_Ar_D} \Bigg)^{l+2} \Bigg|^2
\end{equation}

\noindent while the plasmonic enhancement factor of the FRET process
for dipole arrangements shown in Fig, 1(b) takes the following form

\begin{equation} \label{eq26} 
  \begin{split}
    |A|^2  = & \Bigg|1+\frac{(r_A+r_D)^2}{2a^3}\sum_{l,m}
    \frac{(\epsilon_1-\epsilon_0)}{l \epsilon_1+ (l+1)\epsilon_0}
    \frac{l}{2l+1}
    \times
    \\
    &      \times
    \Bigg( \frac{a^2}{r_Ar_D} \Bigg)^{l+2} 
    \frac{dY_{lm}(\theta, \varphi)}{d\theta}
    \Big|_{\begin{subarray}{l}{\theta=\pi}
        \\ {\varphi=0}\end{subarray}}  \frac{dY^*_{lm}(\theta,
      \varphi)}{d\theta} \Big|_{\begin{subarray}{l}{\theta=0}
        \\ {\varphi=0}\end{subarray}} \Bigg|^2
  \end{split}
\end{equation}

\begin{figure}
  \centering
  \includegraphics[width=0.6\linewidth]{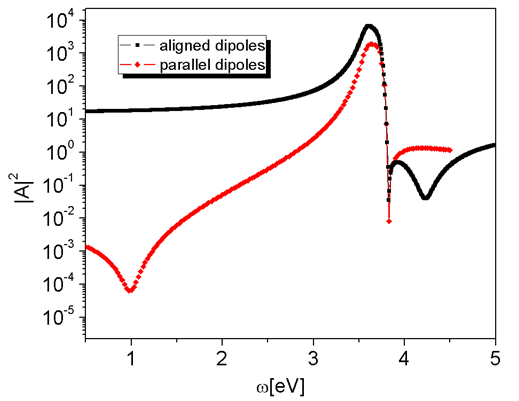}
  \caption{The plasmonic enhancement factor of FRET for both
    arrangement geometries of the donor and the acceptor.  The aligned
    dipole geometry (Fig. 1(a)) is plotted with black line with
    squares and the parallel geometry (Fig. 1 (b)) is plotted with red
    line with diamonds.}
  \label{fig:figure2}
\end{figure}

We have several observations. First, Eq. (25) exhibits only $m=0$ for
the whole sum regarding $l$ and $m$ indices. On the other hand,
careful inspection of Eq. (26) will show that only terms with $m=\pm1$
are different from 0, because only spherical harmonics that has only
$sin(\theta)$ will survive. This can be further shown and Eq. (26) can
be further worked out by using some recurrence relations for the
derivative of associated Legendre polynomials which are part of
spherical harmonics [32]. Also, apparently, due to the factor
$(l+1)^2$ in Eq. (25) and the factor $1/(2l+1)$ in Eq. (26), the
enhancement factor of FRET for aligned dipoles seems to be
higher. Numerical estimations performed below prove our last
observation.

In order to estimate numerically the plasmonic enhancement factor
$|A|^2$ we consider a spherical nanoparticle with 25 nm radius, made
of silver whose dielectric function is described by a Drude model of
the form
$\epsilon_1(\omega)=\varepsilon_0(\varepsilon_\infty-\omega_p^2/(\omega(\omega+i\delta)))$
having the following parameters:
$\varepsilon_\infty=5,\ \omega_p=9.5$, eV, and $\delta=0.15$ eV. The
donor as well as the acceptor molecules is placed at 5 nm from
nanosphere surface.  The computed enhancement factors as function of
frequency (in eV) are shown in Fig. 2 for both dipole
arrangements. The plasmonic enhancement factors of FRET have maximum
values of about 6700 at 3.6 eV for aligned dipole configuration and of
about 1900 at 3.65 eV for parallel dipole configuration. These
calculations are consistent with other recent calculations of FRET
enhancement for these two dipole arrangements [30]. The authors, who
used a fully retarded numerical scheme, have also noticed a shift of
the maximum of FRET enhancement from the dipole resonance to the
quadrupole one, findings that are consistent with our results [30].

\begin{figure}
  \centering
  \includegraphics[width=0.6\linewidth]{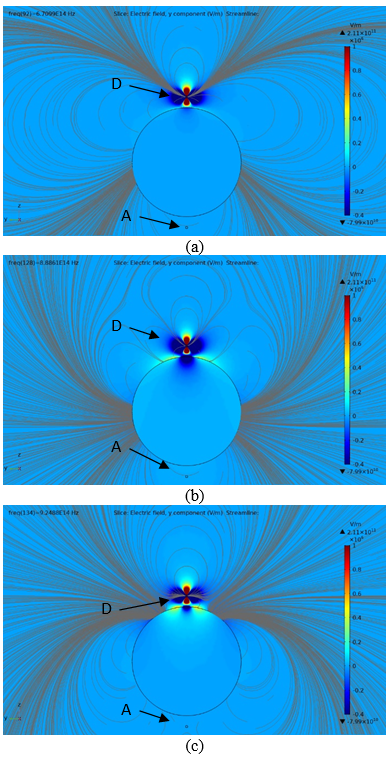}
  \caption{Electric field y-component streamlines at three
    frequencies: (a) 2.78 eV; (b) 3.65 eV; and (c) 3.83 eV. The dipole
    arrangement is parallel.  }
  \label{fig:figure3}
\end{figure}      

To understand the role of the spherical nanoparticle in FRET
enhancement we show the electric field streamlines along y-component,
along the dipole orientations depicted in Fig. 1 (b). The frequencies
are chosen at 2.78 eV (off-resonance), 3.65 eV (on resonance) and at
3.83 eV (at the common dip). In the off-resonance regime, Fig. 3 (a),
the nanosphere behaves as a dipole, but due to the dipole arrangement
it screens the dipole field at acceptor site. The situation is
different from that of the aligned arrangement, where there is an
enhancement due to the sphere induced dipole (see our previous
conference paper [29]). In the on-resonance regime, Fig. 3 (b), the
role of the nanosphere is radically changed. The donor and the
nanosphere become a very large, common, and extended dipole that
enhances the electric field at the acceptor site. However, at the
frequency of the dip (3.83 eV, the same for any dipole arrangement)
the nanosphere effectively screens the donor dipole since some of the
field lines end up on the sphere. We will discuss this issue elsewhere
since it is related to the spectral properties of the electrostatic
operator for spherical shape.
              
To conclude this work, we calculated analytically and analyzed
numerically the enhancement of intermolecular resonance energy
transfer in the presence of a spherical metallic nanoparticle. We
calculated a boundary Green's function that straightforwardly provides
the plasmonic enhancement factor of the FRET and its modal
decomposition.  In contrast to other previous works, our approach
seems to be more direct since it makes use of spectral properties of
the electrostatic operator for spherical geometry. Our numerical
calculations show that at large plasmonic enhancement factors the
donor and the nanoparticle become both a large and extended dipole
that enhances the FRET process.

\textbf{Acknowledgements.}This research was supported by the institutional  CORE-Pro\-gram\-me for 2019 financed by the Romanian Ministry of Research and Innovation.

   \end{document}